Unzipping DNA from the condensed globule state—effects of unraveling

Pui-Man Lam[*] and J.C. Levy
Laboratoire de Physique Theorique de la Matiere Condensee, Universite Paris
7-Denis Diderot, 2 Place Jussieu, 75251 Paris, France

We study theoretically the unzipping of a double stranded DNA from a condensed globule state by an external force. At constant force, we find that the double stranded DNA unzips an at critical force $F_c$ and the number of unzipped monomers M goes as $M \sim (F_c-F)^{-3}$, for both the homogeneous and heterogeneous double stranded DNA sequence. This is different from the case of unzipping from an extended coil state in which the number of unzipped monomers M goes as $M \sim (F_c-F)^{-\chi}$, where the exponent $\chi$ is either 1 or 2 depending on whether the double stranded DNA sequence is homogeneous or heterogeneous respectively. In the case of unzipping at constant extension, we find that for a double stranded DNA with a very large number N of base pairs, the force remains almost constant as a function of the extension, before the unraveling transition, at which the force drops abruptly to zero. Right at the unraveling transition, the number of base pairs remaining in the condensed globule state is still very large and goes as $N^{3/4}$, in agreement with theoretical predictions of the unraveling transition of polymers stretched by an external force.

* On leave from Physics Department, Southern University, Baton Rouge, Louisiana 70813; email: pmlam@grant.phys.subr.edu

# I. INTRODUCTION

In the past decade advances in experimental techniques [1] in atomic force microscopes [2,3], optical tweezers [4,5] and glass microneedles [6,7] have allowed manipulation of single biological molecules, revealing many of their new and unexpected behaviors [8-10]. Of particular interest is the area of DNA molecules where micromanipulation technique involving handles attached to two ends of the molecule were developed, using e.g. biotin-streptavidin linkages and magnetic beads to stretch and twist single DNA molecules by small magnetic field gradients [11,12]. In this way the response of single DNA molecules to external torques [13,14] and its mechanical unzipping in the absence of enzymes [15-19] have been studied. The latter topic has only recently been subjected to theoretical investigations [20-35], while the theoretical studies of thermal denaturation of DNA have even a much longer history [36-47]. Understanding DNA denaturation under conditions of constant force could provide insights into the complicated process by which DNA replicates during bacterial cell division[48].

# II. UNZIPPING DNA FROM A CONDENSED GLOBULE STATE AT CONSTANT FORCE

The unzipping of a double stranded (ds) DNA into two single stranded (ss) DNA by the application of an external force has been studied, for the case when the double stranded DNA is in the extended coil state [13-35]. However, it is known that when the temperature is lowered, the ds-DNA should go from the extended coil state into the condensed globule state [41]. As far as we know, the unzipping of DNA from this condensed globule state has not yet been studied either experimentally or theoretically. Unzipping DNA from the condensed globule state involves both opening up the base pair bonds as well as unraveling the DNA from its condensed state. The unraveling transition when a polymer in the condensed globule state is stretched, in a poor or bad solvent, had been studied theoretically before [49-52]. We will apply this theory to the study of the unzipping of DNA from the condensed globule state. Figure1 shows a picture of a ds-DNA being unzipped from a condensed globule state.

The free energy is given by [49]

$$b\Gamma = -\left(N - \frac{M}{2}\right)a - M \log\left(\frac{\sinh(bFb)}{bFb}\right) - e\sqrt{M} + g'M + g(2N - M)^{2/3} \quad (1)$$

The original ds-DNA consists of N base pairs. Under an external unzipping force F, M/2 base pairs are unzipped, resulting in a ss DNA with M monomers. The first term represents the free energy of the zipped part of the ds-DNA with (N-(M/2)) remaining base pairs. Since the average free energy per base pair of the ds-DNA is approximately $2.5K_BT$, $\alpha \approx 2.5$ [53,54]. The second term represents free energy in the freely jointed chain (FJC) approximation, of the unzipped part with M monomers in ss state. The third term represents the fluctuation contribution from the sequence heterogeneity in the ds-DNA, since the GC base pairs have a higher binding energy, approximately double that of the AT base pairs. The fourth term in (1) represents the penalty in exposing M monomers to the poor solvent and the fifth term represents the surface tension of the globule.

The first three terms alone in (1) would describe usual unzipping of a ds DNA from an extended coil state [35], neglecting self-avoidance. It would give a critical force for unzipping $F_c$ given by the solution of the equation

$$\frac{a}{2} - \log\left(\frac{\sinh(bF_c b)}{bF_c b}\right) = 0 \qquad (2)$$

and the number of unzipped monomers behaves as M~( $F_c$-F)$^{-\chi}$, where the exponent $\chi$ is either 1 or 2 depending on whether the dsDNA sequence is homogeneous or heterogeneous respectively [26-29,34,35].

Equation (1) can be written in the form

$$b\Gamma = -Na + Mt - e\sqrt{M} + g(2N-M)^{2/3} \qquad (3)$$

with

$$t = \frac{a}{2} + g' - \log\left(\frac{\sinh(bFb)}{bFb}\right) \qquad (4)$$

In the limit M,N→∞, for t < 0, equation (3) has no minimum with respect to M. For t > 0, equation (3) has a minimum with respect to M. This shows that t=0 in (4) determines the critical force $F_c$ for unzipping, $t(F_c)=0$. For t > 0, the minimum of the free energy is obtained by setting to zero the derivative of (3) with respect to M. This gives

$$t - \frac{1}{2}eM^{-1/2} - \frac{2}{3}g(2N-M)^{-1/3} = 0 \qquad (5)$$

Mutiplying by $N^{1/3}$, Eqn. (5) can be written as

$$tN^{1/3} - 2^{-3/2}\left(\frac{M}{2N}\right)^{-1/2} N^{-1/6} - \frac{2^{2/3}}{3}g\left(1 - \frac{M}{2N}\right)^{-1/3} = 0 \qquad (6)$$

In the limit M,N→∞, and M/(2N)<1, the second term can be dropped. We have then

$$tN^{1/3} - \frac{2^{2/3}}{3} g \left(1 - \frac{M}{2N}\right)^{-1/3} = 0 \qquad (7)$$

This equation can be written as

$$N = 4\left(1 - \frac{M}{2N}\right)^{-1} \left(\frac{g}{3t}\right)^3 \qquad (8)$$

For $M = \gamma N$, $1 < \gamma < 2$, this equation can be written as

$$\frac{M}{g} = 4\left(1 - \frac{g}{2}\right)^{-1} \left(\frac{g}{3t}\right)^3 \qquad (9)$$

Since $t(F_c)=0$, (9) shows that $M \sim (F_c-F)^{-3}$, for the unzipping of both homogeneous and heterogeneous ds DNA. This is different from unzipping DNA from the extended coil state mainly due to the presence of the surface tension in the globule state.

III. UNZIPPING DNA FROM A CONDENSED GLOBULE STATE AT CONSTANT EXTENSION

In the above we have discussed the unzipping of DNA with a constant force. Experimentally it is possible to generate a constant force through a clever feedback loop mechanism [19]. However, the more usual experimental situation is the unzipping of DNA at constant extension and measuring the resulting force instead. In the case of constant extension, the free energy is given by

$$b\Gamma = -\left(N - \frac{M}{2}\right)\mathbf{a} - \mathbf{e}\sqrt{M} + g'M + g(2N-M)^{2/3} + \frac{1}{b}\int_0^R dx\, l^*\left(\frac{x}{Mb}\right) \quad (10)$$

where $l^*(x)$ is the inverse Langevin function which gives the force at extension x/(Mb) in the FJC model with M monomers each of length b [40]. We have neglected the radius of the globule compared to the extension R. The last term in (10) represents the work done in pulling the ends of the two ss DNA a distance R apart. The force as a function of the extension R is given by

$$F = \frac{\partial \Gamma}{\partial R} = \frac{K_B T}{b} l^*\left(\frac{R}{Mb}\right) \quad (11)$$

The value of M that will minimize the free energy is determined by the solution of the equation

$$\frac{\mathbf{a}}{2} + g' - \frac{\mathbf{e}}{2\sqrt{M}} - \frac{2}{3}g(2N-M)^{-1/3} - \frac{R}{Mb}l^*\left(\frac{R}{Mb}\right) + \frac{1}{M}\int_0^R dx\, l^*\left(\frac{x}{Mb}\right) = 0 \quad (12)$$

For fixed N, the number of base pairs in the ds DNA, taking b=1, and for given values of the parameter $\alpha$, $\varepsilon$, g and g', equation (10) can be solved numerically for M, at fixed values of the extension R, with the restriction that R < Mb. Using this value of M in (8) would give the force at that value of R. In Figure 2 we show the force versus R for $2N=10^7$, with $\alpha=2.5$, g'=1 and g=10, for three different values of the parameter $\varepsilon$, which characterizes the random sequence in ds DNA. From this figure we can see that the parameter $\varepsilon$ has very little effect on the force versus extension curve. It should be pointed out that although ε does not affect the quenched average of the force extension curve in the case of constant extension, it does affect individual realizations which is extremely important in experiments. It should be pointed out that our force extension curve in Fig. 2 starts off at a maximum at the shortest extension (R=0) and then drops to zero. This is very different from those obtained in DNA stretching experiments performed in the coil state using AFM or optical tweezer measurements where the force-extension curves typically start from near zero forces and the force grows until the DNA is fully stretched. The unusual behavior in our force-extension curve is due to surface effect in the globule state. The surface effect is strongest at zero extension when the surface area of the globule is largest and decreases when the DNA starts to unravel and eventually disappears when the DNA is completely unraveled. The force drops to zero if the two single strands are separated. If the two single strands are joined at one end as in the case of a very large RNA hairpin, then the force will increase from zero again as the two single strands are pulled further apart and the usual force-extension curve will be obtained after this point. Perhaps this is one way to experimentally test the surface effect.

For fixed N, equation (12) has solution only for extensions R less than a maximum value $R_{max}$, corresponding to a maximum number of unzipped base pairs $n_{max}$. Beyond $R_{max}$, the force would drop suddenly to zero. The result that the force drops to zero at some critical extension is due to the fact that when the double stranded DNA is completely unzipped, it becomes two separate single strands, not connected to each other at all. Therefore the force must then drop to zero. The interesting thing is that the force is already zero even before the two strands are completely unzipped. This corresponds to the unraveling transition in the case of polymer in the condensed globule state being pulled by an external force [49-52]. There it was shown that for a polymer chain with N monomers, the number of monomers remaining in the globule state at the transition is still large and goes as N- $n_{max}$ ~ $N^{3/4}$ [49]. If the two strands are joined at one end as in the case of a very large, single RNA hairpin, then as the double strand becomes completely unzipped, the force versus extension curve would first drop to zero and then rise again because now the single strand is being stretched by the external force. This would be a more easily observable effect. The stretching within this model is given by the freely joint chain model, which is the main simplification used in this model. In Figure 3 we show in log-log plot N- $n_{max}$ versus N. We find that the slope is about 0.76, close to the value ¾. In Figure 4 we plot the force F versus R for various values of N. From this figure we can see that as N increases, the force stays more and more constant before the transition and the force drops more abruptly to zero at the transition which occurs at larger extensions.

IV. CONCLUSION

We study theoretically the unzipping of a ds DNA from a condensed globule state by an external force. At constant force, we find that the ds DNA unzips an at critical force $F_c$ and the number of unzipped monomers M goes as M~$(F_c-F)^{-3}$, for both the homogeneous and heterogeneous ds DNA sequence. This is different from the case of unzipping from an extended coil state in which the number of unzipped monomers M goes M~$(F_c-F)^{-\chi}$, where the exponent $\chi$ is either 1 or 2 depending on whether the dsDNA sequence is homogeneous or heterogeneous respectively. In the case of unzipping at constant extension, we find that for a ds DNA with a very large number N of base pairs, the force remains almost constant as a function of the extension, before the unraveling transition, at which the force drops abruptly to zero. Right at the unraveling transition, the number of base pairs remaining in the condensed globule state is still very large and goes as $N^{3/4}$, in agreement with theoretical predictions of the unraveling transition of polymers stretched by an external force. Our model of the globular state applies only to very long DNA, which means that experimentally one has to use very long DNA to see this effect. For somewhat shorter DNA the toroidal state is more relevant

Recently, Danilowicz et al [55] separated double stranded λ-DNA by applying a fixed force at constant temperature from 15 to 50 C and measuring the minimum force required to separate the two strands. They found that while part of the force versus temperature phase diagram can be explained using existing models and free energy parameters, i.e. models in good solvent without surface area effect, others deviate significantly. To our knowledge, so far no experiment has be performed on unzipping DNA in a poor solvent. In principle, the experimental technique of Danilowicz can be applied to probe the results of this paper.

Figure Captions

Figure 1: A globule of DNA with N base pairs is unzipped by a force F into two separate ss DNA with M/2 monomers each and end to end extension R.

Figure 2: Force versus extension for three values of the parameter $\varepsilon$, for a DNA with N=$10^7$ base pairs, $\alpha$=2.5, g'=1, g=10.

Figure 3: Number of base pairs remaining in the condensed globule state versus the total number of base pairs N.

Figure 4: Force versus extension curves for DNA with various number of base pairs N, $\alpha$=2.5, g'=1, g=10.

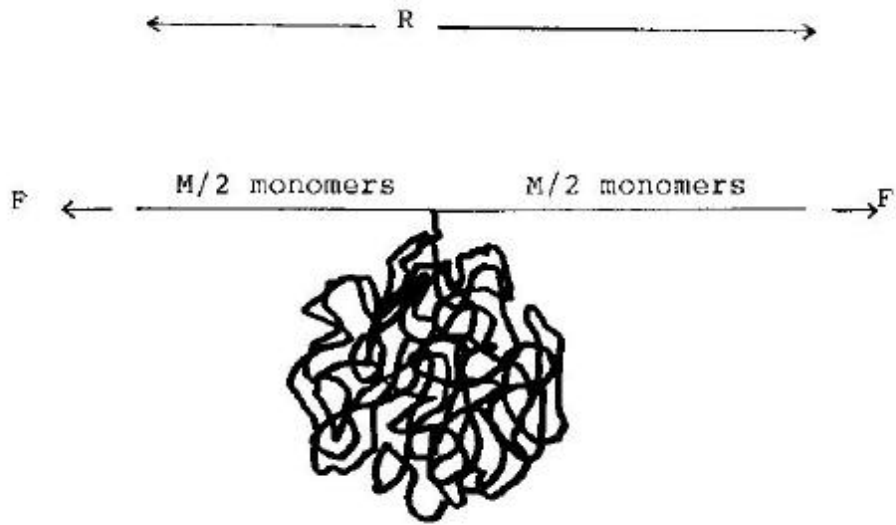

Figure 1: A globule of dsDNA with N base pairs is unzipped by a force F into two separate ss DNA with M/2 monomers each and end-to-end extension R.

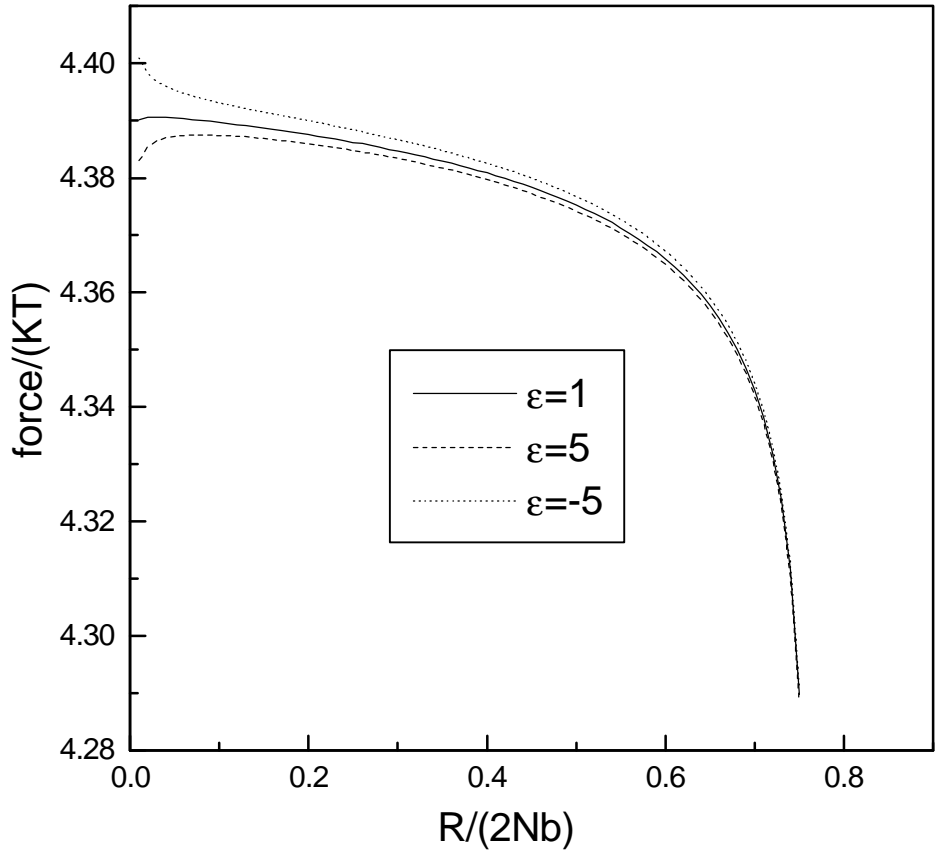

Figure 2: Force versus extension for three values of the parameter $\varepsilon$, for a DNA with $N=10^7$ base pairs, $\alpha=2.5$, $g'=1$, $g=10$.

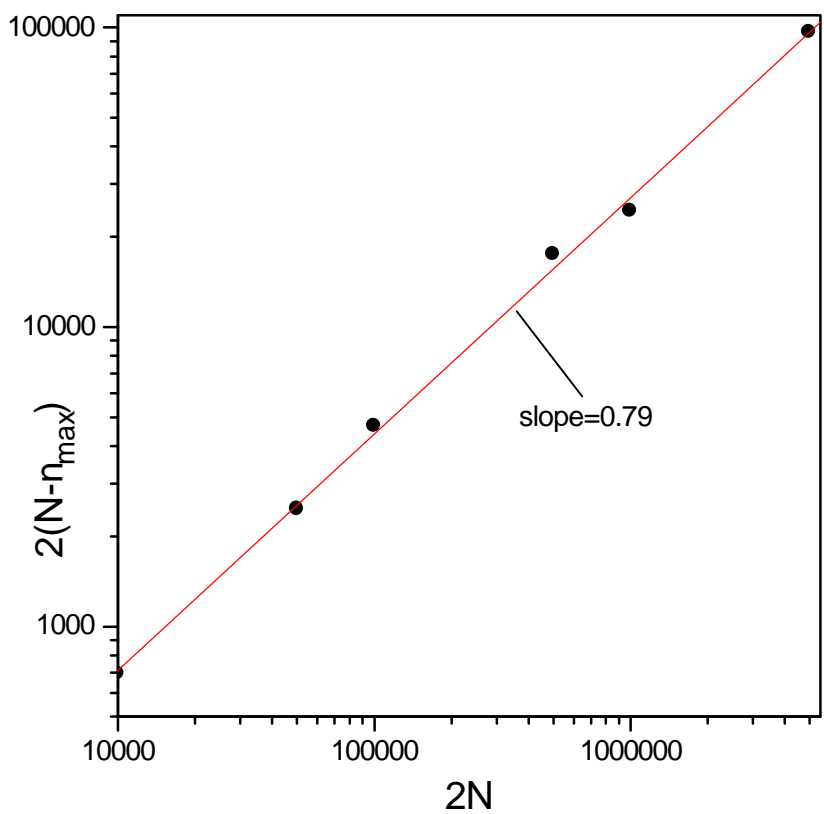

Figure 3: Number of base pairs remaining in the condensed globule state versus the total number of base pairs N.

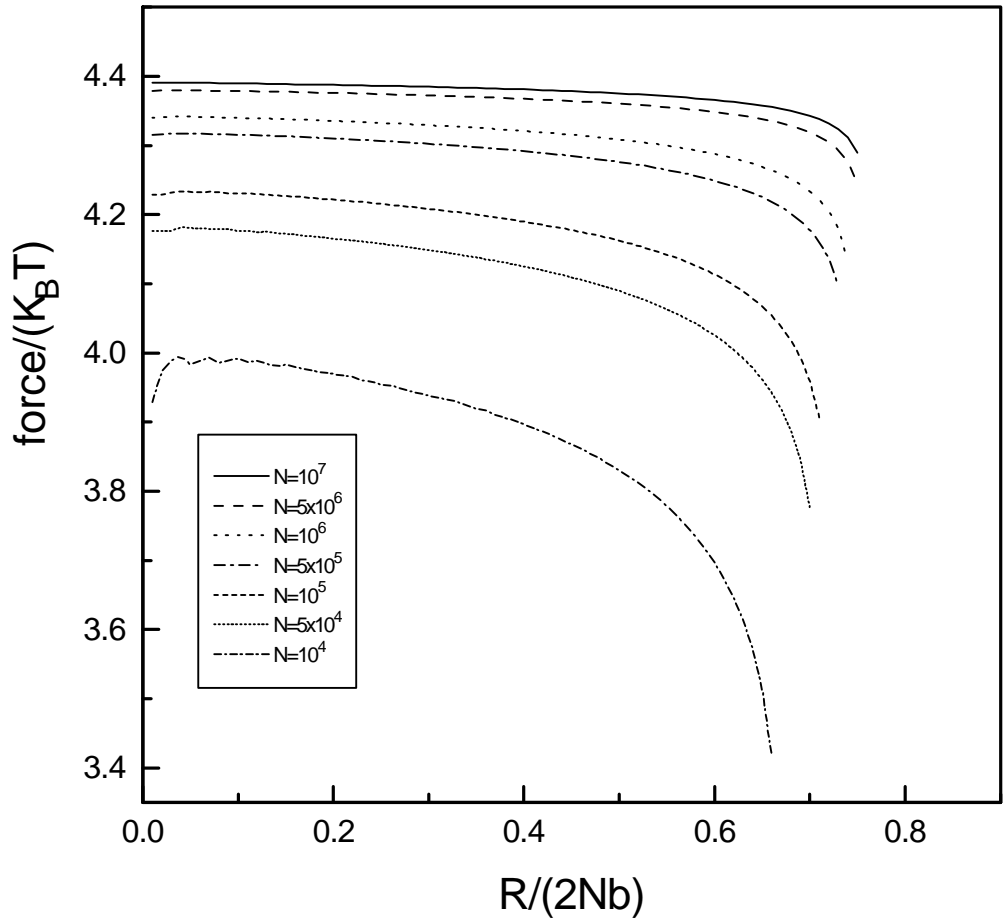

Figure 4: Force versus extension curves for DNA with various number of base pairs N, α=2.5, g'=1, g=10.